\documentclass[twocolumn,showpacs,preprintnumbers,amsmath,amssymb]{revtex4}

\def\undersim#1{\setbox9\hbox{${#1}$}{#1}\kern-\wd9\lower
    2.5pt \hbox{\lower\dp9\hbox to \wd9{\hss $_\sim$\hss}}}
\usepackage{amssymb}
\usepackage{amsmath}
\usepackage{graphicx}
\usepackage{color}

\def\undersim#1{\setbox9\hbox{${#1}$}{#1}\kern-\wd9\lower
    2.5pt \hbox{\lower\dp9\hbox to \wd9{\hss $_\sim$\hss}}}

\def\mv{{\mathbf v}}
\def\mr{{\mathbf r}}

\def\mr{{\mathbf r}}

\def\mk{{\mathbf k}}

\begin{document}

\title{Squeezed correlations of bosons with nonzero widths for expanding sources}

\author{Yong Zhang$^{1}$}
\author{Peng Ru$^{2}$}
\affiliation{\small$^1$School of Mathematics and Physics,
Jiangsu University of Technology, Changzhou, Jiangsu 213001, China\\
$^2$School of Materials and New Energy, South China Normal University, Shanwei 516699, China}


\begin{abstract}
We explore the squeezed back-to-back correlation (SBBC) and investigate how the squeezing effect influences the Hanbury Brown-Twiss (HBT) interferometry
using an expanding Gaussian source with non-zero width.
The SBBC and HBT of $D^0$ and $\phi$ mesons with finite in-medium widths are studied.
The expanding flow of the source may enhance the SBBC strength of $D^0{\bar D}^0$ and $\phi\phi$ in the low momentum region
but suppress the SBBC in the larger momentum region.
The squeezing effect suppresses the influence of flow on the HBT {\color{black}radii},
which is significant for two identical bosons with large pair momentum {\color{black}or with large mass}. Due to the squeezing effect, the relationship between the HBT {\color{black}radii} and the pair momentum exhibits non-flow behavior for $D^0D^0$ pair. Likewise, non-flow behavior also appears in the HBT {\color{black}radii} of $\phi\phi$ with large pair momentum.
This phenomenon may bring new insights for studying the squeezing effect.

\end{abstract}
\maketitle

\section{Introduction}
The interactions between bosons and the source medium in high-energy heavy-ion collisions can cause a squeezing effect,
resulting in a squeezed back-to-back correlation (SBBC) between boson and anti-boson in the particle-emitting source \cite{AsaCso96,AsaCsoGyu99,Padula06,DudPad10,Zhang15a,Zhang-EPJC16,AGY17,XuZhang19,sbbn}.
{\color{black}The squeezing effect and SBBC are connected to the in-medium mass modification of bosons through a Bogoliubov transformation. This
transformation establishes a connection between the creation (annihilation) operators of quasiparticles within the medium and their corresponding free particles \cite{AsaCso96,AsaCsoGyu99,Padula06,DudPad10,Zhang15a,Zhang-EPJC16,AGY17,XuZhang19,sbbn}.}
The investigations of the SBBC may offer a fresh perspective for understanding the dynamic and thermal characteristics of the particle source.

Recently, the SBBC function between a boson and anti-boson with a non-zero width has been formulated and the influences of the in-medium width
on the SBBC functions of $D^0{\bar D}^0$ and $\phi\phi$ {\color{black}have} been studied {\color{black}in} a static homogeneous source \cite{XuZhang19-2}.
However, the particle-emitting sources created in high-energy heavy-ion collisions are both expansive and inhomogeneous.
To create a more realistic model, we extended previous research to the expanding Gaussian source with non-zero width.
The results indicate that the expanding flow of the source may enhance the SBBC strength of $D^0{\bar D}^0$ and $\phi\phi$ in the low momentum region
but suppress the SBBC in the larger momentum region.

Hanbury Brown-Twiss (HBT) interferometry has been extensively used for studying the space-time
properties and coherence of the particle-emitting sources formed in high-energy heavy-ion
collisions \cite{Gyu79,Wongbook,Wie99,Wei00,Csorgo02,Lisa05,hbtl1,hbtl2,hbtl3,hbtl4,hbts1,hbts2,hbts3,hbts4,hbts5,hbts6}.
The correlation functions between two identical bosons can be obtained by taking the ratio of their two-particle momentum spectrum to that of the product of two single boson momentum spectra. The squeezing effect may also affect
the two-particle momentum spectrum and the single boson momentum spectra, respectively \cite{AsaCsoGyu99,Padula06,DudPad10}.
The squeezing effect on the HBT correlation function of two identical kaon was shown utilizing non-relativistic
formulism \cite{DudPad10}. This paper delves deeper into the study of the effect of squeezing on the HBT correlation function{\color{blue}s} of $D^0D^0$
and $\phi$$\phi$ while accounting for nonzero width using relativistic formalism.
{\color{black}In a static source, the HBT radii of two identical bosons show a nearly constant behavior as the pair momentum increases. Conversely, in an expanding source, the HBT radii decrease as the pair momentum increases due to the influence of expanding flow \cite{Lisa05,Heinz-npa96,hbtl1,hbtl4,hbts2,hbts3,hbts4}.}
The squeezing effect suppresses the influence of flow on the HBT {\color{black}radii and causes them to remain almost unchanged or increase with increasing pair momentum. The phenomenon is referred to as the non-flow behavior of the HBT radii in this paper,}
which is significant for two identical bosons with large pair momentum {\color{black}or with large mass}. Due to the squeezing effect, the relationship between the HBT {\color{black}radii} and the pair momentum exhibits non-flow behavior for $D^0D^0$ pair. Likewise, non-flow behavior also appears in the HBT {\color{black}radii} of $\phi\phi$ with large pair momentum.
The recent analyses of experimental data on $D$ and $\phi$ mesons have garnered significant interest \cite{{STAR-PRL21,STAR-PRC20,STAR-PRC19D,CMS-PRL18D,CMS-PLB18D,
ALICE-JHEP22,ALICE-EPJC20,ALICE-JHEP18D,
ALICE-JHEP16D,ALICE-JHEP15D,ALICE-PRC14D,ALICE-PRL13D,ALICE-EPJC23,ALICE-PRC22,ALICE-EPJC21,ALICE-EPJC18p,STAR-PRL16p,
STAR-PRC16p,PHENIX-PRC16p,ALICE-PRC15p,PHENIX-PRC11p,STAR-PLB09p,STAR-PRC09p,
STAR-PRL07p,PHENIX-PRL07p,PHENIX-PRC05p,NA50-PLB03p,STAR-PRC02p,NA50-PLB00p}}.
This is due to the presence of a charm or strange quark, which is believed to undergo the complete
evolution of the quark-gluon plasma (QGP) formed in high-energy heavy-ion collisions.
On the other hand, the in-medium modifications of masses and widths of $D$ and $\phi$ mesons were expected to exist in
the particle-emitting sources formed in high-energy heavy-ion collisions \cite{Dm,dm2,dm3,dm4,pm,pm1,pm2}.
Thus, the study of the squeezing effect on $D$ and $\phi$ mesons is meaningful in high-energy heavy-ion collisions.

In the next section, we generalize the formulas of the correlation between two bosons with nonzero width for local-equilibrium expanding sources
based on the work in Ref. \cite{AsaCsoGyu99,XuZhang19-2}.
Then, we show the effect of flow on the SBBC of $D^0{\bar D}^0$ and $\phi\phi$ in Section III, and the squeezing effect on the HBT interferometry
of $D^0D^0$ and $\phi\phi$ is also shown in Section III. Finally, summary and discussion are given in Section IV.

\section{Formulas}
Denote $a^\dagger_\mk (a_\mk)$ as the creation
(annihilation) operators of the boson in vacuum with momentum $\mk$, mass $m$, and width $\Gamma_0$.
Similarly, denote $b^\dagger_\mk (b_\mk)$ as the creation
(annihilation) operators of the corresponding quasi-particle
with momentum $\mk$, modified mass $m_{\!*}$, and modified width $\Gamma$ in the medium.
They are related by the transformation \cite{XuZhang19-2}
\begin{equation}
e^{i\omega_\mk t}a^{\dag}_\mk=c^*_\mk e^{i\Omega_\mk t}b^{\dag}_\mk
+s_{-\mk} e^{-i\Omega_\mk t}b_{-\mk},
\label{B-transp1}
\end{equation}
\begin{equation}
e^{-i\omega_\mk t}a_\mk=c_\mk e^{-i\Omega_\mk t}b_\mk +s^*_{-\mk}
e^{i\Omega_\mk t} b^{\dag}_{-\mk},
\label{B-transp}
\end{equation}
\begin{equation}
c_{\pm\mk}=\frac{\cosh r +i\cosh f}{\sqrt{2}},~~
s_{\pm\mk}=\frac{\sinh r +i\sinh f}{\sqrt{2}},
\label{cpsp}
\end{equation}
where $r$ and $f$ are
\begin{equation}
r=\frac{1}{2}\ln\left[\frac{|\omega_\mk|(1- \sin(\Theta-\theta))}
{|\Omega_\mk|\cos(\Theta-\theta)}\right],
\label{rn}
\end{equation}
\begin{equation}
f=\frac{1}{2}\ln\left[\frac{|\omega_\mk|(1+ \sin(\Theta-\theta))}
{|\Omega_\mk|\cos(\Theta-\theta)}\right],
\label{fn}
\end{equation}
here $\omega_\mk$ and $\Omega_\mk$ are the energy of bosons in vacuum and in medium, respectively.
\begin{equation}
\omega_\mk=\sqrt{\mk^2+(m_0-i\Gamma_0/2)^2}= |\omega_\mk| e^{i\theta},
\end{equation}
\begin{equation}
\Omega_\mk =\sqrt{\mk^2+(m_*-i\Gamma/2)^2}= |\Omega_\mk|\,
e^{i\Theta},
\end{equation}
where
\begin{equation}
|\omega_\mk|=\left\{\bigg[\mk^2\!+\!m_0^2\!-\!\frac{\Gamma_0^2}{4}
\bigg]^2\!+\!m_0^2\Gamma_0^2\right\}^{\!{1/4}},
\end{equation}
\begin{equation}
|\Omega_\mk|=\left\{\bigg[\mk^2\!+\!m_*^2\!-\!\frac{\Gamma^2}{4}
\bigg]^2\!+\!m_*^2\Gamma^2\right\}^{\!{1/4}},
\end{equation}
\begin{equation}
\theta=\frac{1}{2}\tan^{-1}\left[\frac{-m_0\Gamma_0}
{\mk^2\!+\!m_0^2\!-\!\Gamma_0^2/4}\right],
\end{equation}
\begin{equation}
\Theta=\frac{1}{2}\tan^{-1}\left[\frac{-m_*\Gamma}
{\mk^2\!+\!m_*^2\!-\!\Gamma^2/4}\right].
\end{equation}
It is necessary to mention that the above transformation can only be regarded as an approximation if the
imaginary part of $-2\Omega_\mk$~[\,${\rm Im}(-2\Omega_\mk)\!\sim\!m_0
\Gamma/\omega_\mk\!\sim\!\Gamma$ for~$\mk^2\!<\!m_0^2$\,] is not considerably large \cite{XuZhang19-2}.

The correlation function of the two bosons with momenta $\mk_1$ and $\mk_2$ is
defined as \cite{AsaCso96,AsaCsoGyu99}
\begin{equation}
\label{BBCf}
C(\mk_1,\mk_2) = 1 + \frac{|G_c(\mk_1,\mk_2)|^2+|G_s(\mk_1,\mk_2)|^2}{G_c(\mk_1,\mk_1) G_c(\mk_2,
\mk_2)},
\end{equation}
where $G_c(\mk_1,\mk_2)$ and $G_s(\mk_1,\mk_2)$ are the chaotic and squeezed amplitudes
\begin{equation}
G_c(\mk_1,\mk_2)=\sqrt{\omega_{\mk_1}\omega_{\mk_2}}\,\langle a^{\dagger}_{\mk_1}
a_{\mk_2}\rangle ,
\end{equation}
\begin{equation}
G_s(\mk_1,\mk_2)=\sqrt{\omega_{\mk_1}\omega_{\mk_2}}\,\langle a_{\mk_1}
a_{\mk_2}\rangle.
\end{equation}
For further discussion, the labeling convention for particles and antiparticles is as follows:
$"+"$ denotes particles, $"-"$ denotes antiparticles $($unless the antiparticle of the particle is itself,
in which case $"0"$ is used for both$)$. The correlation functions of the two bosons for pairs of $(0\,0)$, $(+\,-)$,
and $(+\,+)$ become \cite{AsaCsoGyu99}
\begin{equation}
\label{BBCf1}
C^{00}(\mk_1,\mk_2) = 1 + \frac{|G_c(\mk_1,\mk_2)|^2+|G_s(\mk_1,\mk_2)|^2}{G_c(\mk_1,\mk_1) G_c(\mk_2,
\mk_2)},
\end{equation}
\begin{equation}
\label{BBCf2}
C^{+-}(\mk_1,\mk_2) = 1 + \frac{|G_s(\mk_1,\mk_2)|^2}{G_c(\mk_1,\mk_1) G_c(\mk_2,
\mk_2)},
\end{equation}
\begin{equation}
\label{BBCf3}
C^{++}(\mk_1,\mk_2) = 1 + \frac{|G_c(\mk_1,\mk_2)|^2}{G_c(\mk_1,\mk_1) G_c(\mk_2,
\mk_2)},
\end{equation}
Eq.~\ref{BBCf1} describes the correlation function of two particles where the antiparticle is itself.
Eq.~\ref{BBCf2} describes the correlation function of boson and anti-boson, and it is also known as SBBC when $\mk_1=\,-\mk_2$.
Eq.~\ref{BBCf3} describes the correlation function of two identical bosons, and it is also called HBT interferometry.

For local-equilibrium expanding sources, $G_c(\mk_1,\mk_2)$ and $G_s(\mk_1,\mk_2)$
can be expressed as \cite{AsaCsoGyu99,Padula06,DudPad10,Zhang15a,Zhang-EPJC16,AGY17,XuZhang19}
\begin{eqnarray}
&& \hspace*{-7mm} G_c({\mk_1},{\mk_2})\!=\!\int \frac{d^4\sigma_{\mu}(x)}{(2\pi)^3}
K^\mu_{1,2} e^{i\,q_{1,2}\cdot x}\,\! \Bigl\{|c'_{\mk'_1,\mk'_2}|^2\,n'_{\mk'_1,
\mk'_2}\nonumber \\
&& \hspace*{11mm} +\,|s'_{-\mk'_1,-\mk'_2}|^2\,[\,n'_{-\mk'_1,-\mk'_2}+1]\Bigr\},
\end{eqnarray}
\begin{eqnarray}
&& \hspace*{-7mm} G_s({\mk_1},{\mk_2})\!=\!\int \frac{d^4\sigma_{\mu}(x)}{(2\pi)^3}
K^\mu_{1,2}e^{2 i\,K_{1,2}\cdot x}\!\Bigl\{s'^*_{-\mk'_1,\mk'_2}c'_{\mk'_2,-\mk'_1}
\nonumber \\
&& \hspace*{3mm}\times n'_{-\mk'_1,\mk'_2}+\,c'_{\mk'_1,-\mk'_2}\,s'^*_{-\mk'_2,
\mk'_1}\,[\,n'_{\mk'_1,-\mk'_2} + 1] \Bigr\},
\end{eqnarray}
where $\mk'_i$
is the local-frame momentum corresponding to $\mk_i~(i=1,2)$.  The other local variables
are:

\begin{equation}
c'_{\pm\mk'_1,\pm\mk'_2}=\frac{\cosh r'_{\pm\mk'_1,\pm\mk'_2} +i\cosh f'_{\pm\mk'_1,\pm\mk'_2}}{\sqrt{2}},
\end{equation}
\begin{equation}
s'_{\pm\mk'_1,\pm\mk'_2}=\frac{\sinh r'_{\pm\mk'_1,\pm\mk'_2} +i\sinh f'_{\pm\mk'_1,\pm\mk'_2}}{\sqrt{2}},
\end{equation}
\begin{eqnarray}
&&\hspace*{-4mm}r'_{\pm\mk'_1,\pm\mk'_2}=\frac{1}{2}\ln\left[\frac{|\omega'_{\mk'_1,\mk'_2}|(1- \sin(\Theta'_{\mk'_1,\mk'_2}-\theta'_{\mk'_1,\mk'_2}))}
{|\Omega'_{\mk'_1,\mk'_2}|\cos(\Theta'_{\mk'_1,\mk'_2}-\theta'_{\mk'_1,\mk'_2})}\right],\nonumber\\
&&
\end{eqnarray}

\begin{eqnarray}
&&\hspace*{-4mm}f'_{\pm\mk'_1,\pm\mk'_2}=\frac{1}{2}\ln\left[\frac{|\omega'_{\mk'_1,\mk'_2}|(1+ \sin(\Theta'_{\mk'_1,\mk'_2}-\theta'_{\mk'_1,\mk'_2}))}
{|\Omega'_{\mk'_1,\mk'_2}|\cos(\Theta'_{\mk'_1,\mk'_2}-\theta'_{\mk'_1,\mk'_2})}\right],\nonumber\\
&&
\end{eqnarray}

\begin{eqnarray}
&&\hspace*{-7mm}\omega'_{\mk'_1,\mk'_2}(x)=\frac{1}{2}\left[\omega'_{\mk'_1}(x)+\omega'_{\mk'_2}(x)\right]\nonumber\\
&&\hspace*{8.3mm}= |\omega'_{\mk'_1,\mk'_2}| e^{i\theta'_{\mk'_1,\mk'_2}},
\end{eqnarray}
\begin{eqnarray}
&&\hspace*{-7mm}\Omega'_{\mk'_1,\mk'_2}(x)=\frac{1}{2}\left[\Omega'_{\mk'_1}(x)+\Omega'_{\mk'_2}(x)\right]\nonumber\\
&&\hspace*{8.3mm}= |\Omega'_{\mk'_1,\mk'_2}| e^{i\Theta'_{\mk'_1,\mk'_2}},
\end{eqnarray}

\begin{eqnarray}
&&\hspace*{-7mm}\omega'_{\mk'_i}(x)=\sqrt{\mk'^2_i(x)+(m_0-i\Gamma_0/2)^2}=k^{\mu}_i u_{\mu}(x)\nonumber\\
&&\hspace*{3.2mm}=\gamma_\mv\,[\,\omega_{\mk_i}-\mk_i\cdot\mv(x)\,],
\end{eqnarray}
\begin{eqnarray}
\label{Omp}
&&\hspace*{-7mm}\Omega'_{\mk'_i}(x)=\sqrt{\mk'^2_i(x)+(m_*-i\Gamma/2)^2}\nonumber\\
&&\hspace*{3mm}=\sqrt{[k^{\mu}_i u_{\mu}(x)]^2-(m_0-i\Gamma_0/2)^2+(m_*-i\Gamma/2)^2},\nonumber\\
\end{eqnarray}
\begin{eqnarray}
\label{nkk}
&&\hspace*{-8mm}n'_{\pm\mk'_1,\pm\mk'_2}=\exp\left\{-\left[\Omega'_{\mk'_1,\mk'_2}(x)-\mu_{1,2}(x)\right]\bigg/T(x)\right\}.\nonumber\\
\end{eqnarray}

In this paper, the spatial distribution of the source is taken as $\rho(x)=C e^{-\mr^2/(2R^2)}\, \theta(r-2R)$.
$R$ and $C$ are the source radius and the normalization constant, respectively. $\mv = \frac{u}{2R}\mr$ is used to represent the expanding velocity of the source and $u$
is a parameter. The time distribution of source is assumed to be the typical exponential decay \cite{AsaCsoGyu99,Padula06,DudPad10}
\begin{equation}
F(t) = \frac{\theta(t-t_0)}{\Delta t} \;e^{-(t-t_0)/\Delta t},
\end{equation}
where $\Delta t$ is a parameter and the time distribution of the source widens as the parameter $\Delta t$ increases.

\section{Results}
{\color{black}In this section, all results are based on Monte Carlo simulation calculations using a spherically symmetric Gaussian expanding source. The parameter $R$, which is chosen to be 7 fm \cite{Padula06,YZHANG_CPC15}, determines the spatial radius of the source. The radial expanding flow of the source is determined by the parameter $u$, and when $u$ = 0, it represents a static source. The time distribution of the source is independent of the spatial distribution and is described by the parameter $\Delta t$. In all figures except for Fig. \ref{cnihedt}, $\Delta t$ is set to 2 fm/$c$ \cite{Padula06,YZHANG_CPC15}.

The freeze-out temperatures of
$D^0$ meson and $\phi$ meson are taken as 150 MeV and 140 MeV, respectively \cite{Padula06,AGY17}. The mass and width of $D^0$ meson in a vacuum, denoted by $m_0$ and $\Gamma_0$ respectively,
are taken as 1864.86 MeV and 0 MeV respectively, and the $m_0$ and $\Gamma_0$ of $\phi$ meson are taken as 1019.46 MeV and 4.26 MeV respectively \cite{PDG-PRD12,PDG-PRD18}.
The modified mass and width in the medium are represented by $m_*$ and $\Gamma$, respectively. The in-medium mass-shift is denoted as $\delta m$, and $\delta m = m_*-m_0 $.
When $\delta m =$ 0 and $\Gamma$ = $\Gamma_0$, the mass and width of bosons in the medium are identical to those in vacuum. This indicates the squeezing effect is not considered in the calculation.}

\subsection{Effect of flow on SBBC}
\begin{figure}[htbp]
\vspace*{0mm}
\includegraphics[scale=0.6]{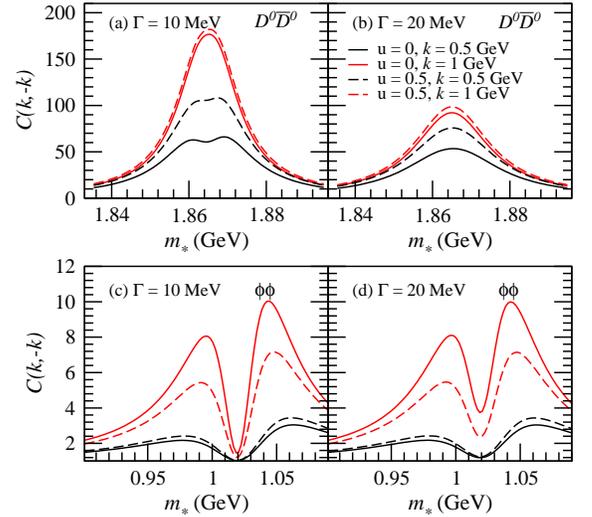}
\caption{(Color online) SBBC results of $D^0{\bar D}^0$ (top panels) and $\phi\phi$ (bottom panels) with respect to modified mass $m_*$ for flow parameter $u$ = 0 and 0.5.  }
\label{ckms}
\end{figure}

In Fig. \ref{ckms}, SBBC results of $D^0{\bar D}^0$ (top panels) and $\phi\phi$ (bottom panels) with respect to modified mass $m_*$ for flow parameter $u$ = 0 and 0.5
are shown.
From Fig. \ref{ckms}, it can be seen that a change of width can lead to a SBBC signal even for {\color{black}$\delta m =$ 0}, especially for $D^0{\bar D}^0$ pair.
The SBBC function of $D^0{\bar D}^0$ is enhanced by the expanding flow for $k = 0.5$ GeV and 1 GeV.
The expanding flow enhances the SBBC signal of $\phi\phi$ for $k = 0.5$ GeV but suppresses the SBBC signal of $\phi\phi$ for $k = 1$ GeV.

\begin{figure}[htbp]
\vspace*{0mm}
\includegraphics[scale=0.61]{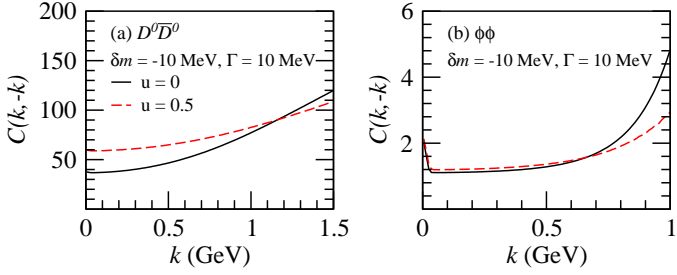}
\caption{(Color online) SBBC results of $D^0{\bar D}^0$ [(a)] and the correlation of $\phi\phi$ [(b)] with respect to momentum $k$ for flow parameter $u$ = 0 and 0.5.
Here, {\color{black}$\delta m$} = -10 MeV and $\Gamma =$ 10 MeV.  }
\label{ckk}
\end{figure}
In Fig. \ref{ckk}, we plot the SBBC results of $D^0{\bar D}^0$ [(a)] and the correlation of $\phi\phi$ [(b)] with respect to momentum $k$ for flow parameter $u$ = 0 and 0.5. Here, $\delta m$ = -10 MeV and $\Gamma =$ 10 MeV. Flow has opposite effects on SBBC of $D^0{\bar D}^0$ depending on the value of momentum $k$, it enhances SBBC when
$k$ is less than 1.2 GeV but suppresses it when $k$ is greater than 1.2 GeV.
For $\phi\phi$ pair, flow suppresses the SBBC signal for $k > 0.7$ GeV but slightly enhances the SBBC signal for $k < 0.7$ GeV. This phenomenon is similar
to the situation without considering the width \cite{Padula06,YZHANG_CPC15}. It is necessary to mention that
Eq. \ref{BBCf1} is used for calculating the correlation of $\phi\phi$. Since the correlations of $\phi\phi$ were shown for $k_1 = -k_2$ in Fig. \ref{ckms} and
Fig. \ref{ckk}, the square of chaotic amplitude $|G_c(\mk_1,\mk_2)|^2$ is almost $0$ and does not contribute to the value of the correlation function
of $\phi\phi$ unless the momentum $k$ is very small (see Fig. \ref{ckk} (b)).

\subsection{Squeezing effect on HBT interferometry}
\begin{figure}[htbp]
\vspace*{0mm}
\includegraphics[scale=0.6]{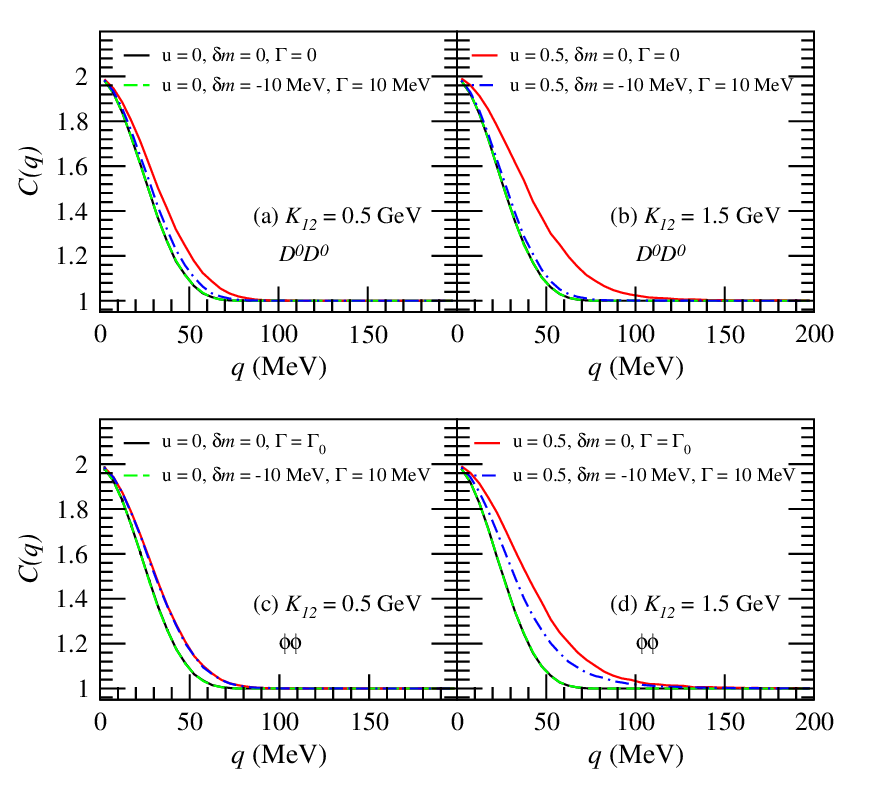}
\caption{(Color online) HBT results of $D^0D^0$ (top panels) and the correlation of $\phi\phi$ (bottom panels) with respect to relative momentum $q$.
Here, $K_{12}$ is the pair momentum and $2K_{12}=|\mk_1+\mk_2|$. }
\label{chbt}
\end{figure}
In Fig. \ref{chbt}, we plot the HBT results of $D^0D^0$ (top panels) and the correlation of $\phi\phi$ (bottom panels) with respect to relative momentum $q$.
Here, the angle between the momentum of two particles is taken as zero, and the square of
squeezed amplitude $|G_s(\mk_1,\mk_2)|^2$ is almost $0$ and does not contribute to the value of the correlation function of $\phi\phi$.
The HBT correlation of $D^0D^0$ or $\phi\phi$ is not affected by
the squeezing effect for static sources. Comparing the results for $u=0$ and $u=0.5$, it can be concluded that the HBT curves are widened by the expanding flow and
the HBT {\color{black}radii} also decrease. The squeezing effect suppresses the widening effect of flow on the HBT curve of $D^0D^0$ for $K_{12}$ = 0.5 GeV and 1{\color{black}.5} GeV.
The squeezing effect does not affect the HBT result of $\phi\phi$ for $K_{12}$ = 0.5 GeV. The widening effect of flow is suppressed by the squeezing effect
for $K_{12}$ = 1.5 GeV for $\phi\phi$ pair. The phenomenon of $\phi$ meson is similar to the non-relativistic results of kaon \cite{DudPad10}.

\begin{figure}[htbp]
\vspace*{0mm}
\includegraphics[scale=0.6]{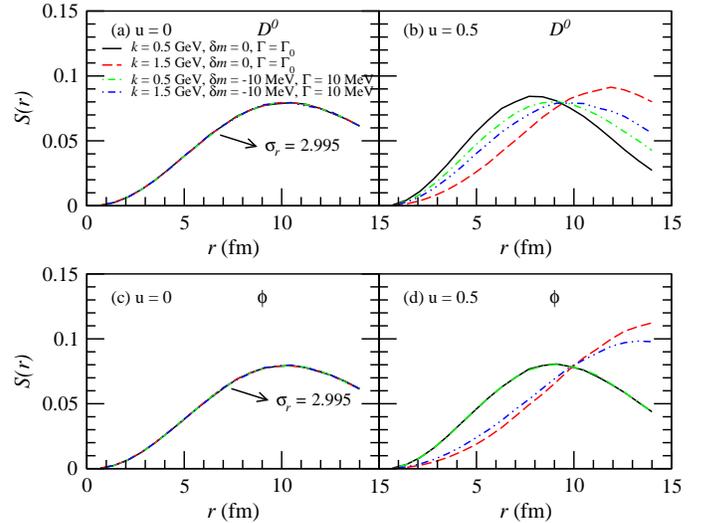}
\caption{(Color online) Normalized distributions of radial coordinates of the sources of $D$ (top panels) and $\phi$ (bottom panels) for $k$ = 0.5 GeV and 1.5 GeV. }
\label{fr}
\end{figure}

\begin{table*}
	\caption{\label{tab1}Standard deviation $\sigma_r$ of emission source of $D$ and $\phi$ for $u = 0.5$. When squeezing on, the values of
$\delta$ and $\Gamma$ are -10 MeV and 10 MeV, respectively.}
	\begin{ruledtabular}
		\begin{tabular}{ccccc}
			&\multicolumn{2}{c}{$D$}&\multicolumn{2}{c}{$\phi$}\\
			Momentum~&~Squeezing\,\,off&Squeezing\,\,on\,&~Squeezing\,\,off&~Squeezing\,\,on\\
            \hline
			0.5 GeV~&~2.939~&~2.998~&~2.990&~2.988\\
            1.5 GeV ~&~2.833~&~2.993~&~2.729&~2.842\\
		\end{tabular}
	\end{ruledtabular}
\end{table*}

We show in Fig. \ref{fr} the normalized distributions of radial coordinates of the sources of $D$ and $\phi$ for $k$ = 0.5 GeV and 1.5 GeV.
For static sources, the squeezing effect does
not affect the distributions of radial coordinates of the sources of $D$ and $\phi$.
Generally, particles with low momentum tend to be generated in greater numbers at lower flow velocities, whereas locations with higher flow velocities tend to generate more particles with high momentum. In the model of this paper, the expanding flow increases as the radial position increases. Hence, the expanding flow leads to a shift in the source distributions for a momentum of 0.5 GeV towards smaller radial positions.
This can be observed by comparing the black solid lines in Figures (a) and (b), or in Figures (c) and (d).
Similarly, the expanding flow causes a shift in the source distributions for a momentum of 1.5 GeV towards larger radial positions.
This can be observed by comparing the red dashed lines in Figures (a) and (b), or in Figures (c) and (d).
For expanding sources, the squeezing effect suppresses the influence of flow on spatial distribution of $D$.
Similarly, it also mitigates the influence of flow on the spatial distribution of $\phi$ for a momentum of 1.5 GeV.
However, it is important to note that the squeezing effect does not alter the influence of flow on the spatial distribution of $\phi$ for a momentum of 0.5 GeV.
The above phenomenon can be attributed to the fact that the response of $|s'_{-\mk'_1,-\mk'_2}|^2$ to flow velocity is significantly lower compared to
that of $n'_{-\mk'_1,-\mk'_2}$ in $G_c({\mk_1},{\mk_2})$ in Eq.\,18 \cite{Zhang20ij}, since the $G_c({\mk_1},{\mk_2})$ is the single particle momentum distribution for $\mk_1 \,= \,\mk_2$.
In Table \ref{tab1}, we show the standard deviation $\sigma_r$ of emission source of $D$ and $\phi$ for $u = 0.5$,
where $\sigma_r = \sqrt{\frac{1}{N}\sum\limits_{i=1}^N(r_{i}-\bar{r})^2}$ and it can qualitatively describe the relative spatial distribution of the source.
Here $N$ is the total number of $D$ or $\phi$ emitted from the source, $r_i$ is the radial coordinate of the particle denoted by $i$ and
$\bar{r}$ is the average radial coordinate. When squeezing off, the value of $\sigma_r$ at $u = 0.5$ is less than that at $u = 0$
(\,$\sigma_r$ for $u = 0$ is shown in Fig. \ref{fr} (a) and (c).\,),
which is more obvious for $k = 1.5$ GeV.
For expanding source ($u = 0.5$), the squeezing effect suppresses the influence of flow on the spatial distribution of the sources
and leads to a larger $\sigma_r$.
From the results of Fig. \ref{chbt}\,--\,\ref{fr} and Table \ref{tab1},
it can be concluded that the expanding flow causes the HBT correlation curve to appear wider by reducing the spatial distribution width of the particle source.
However, the squeezing effect suppresses the influence of flow on the spatial distribution of the sources and leads to a narrower HBT correlation curve compared
to the case without the squeezing effect.

To conduct a more in-depth analysis of the influence of the squeezing effect on the HBT correlation, we extract the one-dimension HBT radius $R_H$ by fitting the
correlation function with the parametrized formula
\begin{eqnarray}
\label{nihe}
C(q)=1+\lambda e^{-q^2R_H^2}.
\end{eqnarray}
{\color{black}It is worth  mentioning that that $D^0$ and $\phi$ mesons are electrically neutral, rendering them unaffected by the Coulomb effect. Consequently, these mesons serve as ideal probes for investigating the squeezing effect compared to charged bosons. Additionally, when performing HBT radii fitting for $D^0$ and $\phi$ mesons, there is no need to account for the Coulomb effect \cite{Lisa05,hbts1}.}

\begin{figure}[]
\vspace*{0mm}
\includegraphics[scale=0.6]{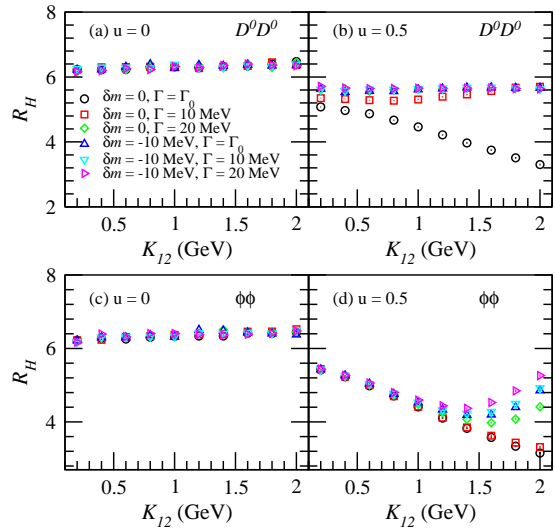}
\caption{(Color online) HBT {\color{black}radii} of $D^0D^0$ (top panels) and $\phi\phi$ (bottom panels) for $u=0$ and $u=0.5$.}
\label{cnihe}
\end{figure}
In Fig. \ref{cnihe}, we plot the HBT {\color{black}radii} of $D^0D^0$ (top panels) and $\phi\phi$ (bottom panels) for $u=0$ and $u=0.5$. For static sources,
the HBT {\color{black}radii} remain almost unchanged as pair momentum $K_{12}$ increases, and it is not affected by the squeezing effect.
The reduction of the HBT {\color{black}radii} are caused by the expanding flow, and this phenomenon intensifies as the pair momentum $K_{12}$ increases.
Thus, the HBT {\color{black}radii} decrease as the pair momentum $K_{12}$ increases for the expanding sources without squeezing effect.
For expanding sources with squeezing effect, the HBT {\color{black}radii} of $D^0D^0$ pair remain almost unchanged as pair momentum $K_{12}$ increases,
and this phenomenon exits even for $\delta m = 0$ and $\Gamma \neq \Gamma_0$.
For expanding sources, the squeezing effect does not influence the HBT {\color{black}radii} of $\phi\phi$ when the pair momentum is small.
However, it can result in larger HBT {\color{black}radii} of $\phi\phi$ for greater pair momentum,
and as the $\delta m$ or $\Gamma$ increases, this phenomenon becomes more noticeable.
It should be noted that even though the $\sigma_r$ of the $D$ source is slightly greater for $\delta$ = -10 MeV, $\Gamma$ = 10 MeV, and $u = 0.5$
compared to $u = 0$ at $k = 0.5$ GeV,
the HBT radius of $D^0D^0$ actually less than the HBT radius for $u = 0$.
This occurs because the fitting analysis relies on a Gaussian form, where the flow and squeezing effect affect the width of the source distribution while also
result in deviations from the Gaussian distribution of the source. Moreover, the non-Gaussian characteristics of the source introduce certain impacts on the fitting results.

\begin{figure}[htbp]
\vspace*{2mm}
\includegraphics[scale=0.65]{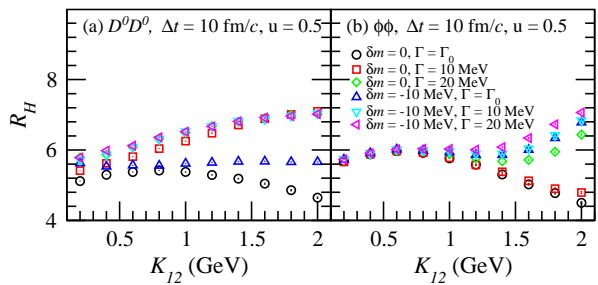}
\caption{(Color online) HBT {\color{black}radii} of $D^0D^0$ [(a)] and $\phi\phi$ [(b)] for $\Delta t = 10$ fm/$c$ and $u=0.5$.}
\label{cnihedt}
\end{figure}

In Fig. \ref{cnihedt}, we plot the HBT {\color{black}radii} of $D^0D^0$ [(a)] and $\phi\phi$ [(b)] for $\Delta t = 10$ fm/$c$ and $u=0.5$.
For expanding sources without squeezing effect, the HBT {\color{black}radii} exhibit a slight increase as the pair momentum increases for small values of $K_{12}$.
For $D^0D^0$ pair, the squeezing effect leads to an increase in the HBT {\color{black}radii} and causes the HBT {\color{black}radii} to increase with the increasing pair momentum $K_{12}$.
The squeezing effect has no impact on the HBT {\color{black}radii} for small pair momentum in the $\phi\phi$ pair, however,
it results in an increase in the HBT {\color{black}radii} for large pair momentum and leads to the HBT {\color{black}radii} to increase with the increasing pair momentum $K_{12}$.
According to the above results, the squeezing effect diminishes the impact of flow on the HBT {\color{black}radii}, particularly for larger pair momentum values.
As the collision energy increases, the bosons may experience a more pronounced medium effect, and the time distribution of the source
becomes wider. The SBBC is very sensitive to the time distribution of the source and may be suppressed to no signal for a wide temporal
distribution \cite{Padula06,DudPad10,Zhang15a,Zhang-EPJC16}.
Thus, the above phenomenon of HBT {\color{black}radii} caused by squeezing effect may provide evidence of the existence of the squeezing effect for the sources
with wide temporal distribution.

{\color{black}The results in this subsection indicate that the squeezing effect suppresses the influence of flow on the spatial distribution of bosons, resulting in a non-flow behavior observed in the relationship between HBT radii and
the pair momentum.
In the calculation, the time distribution of the source is assumed to be independent of the spatial distribution. However, the temporal distribution is correlated with the spatial distribution \cite{Zhang15a,Zhang-EPJC16,spc}. Consequently, the squeezing effect not only affects the spatial distribution but may also influence the time distribution of the source. Unfortunately, the model
employed in this paper is unable to capture the influence of the squeezing effect on the time distribution.}

\section{Summary and discussion}
The interactions between bosons and the source medium in high-energy heavy-ion collisions can cause a squeezing effect,
resulting in a SBBC between boson and anti-boson in the particle-emitting source.
In this paper, we explore the SBBC and investigate how the squeezing effect influences the HBT using an expanding Gaussian source with non-zero width.
The results indicate that the expanding flow of the source may enhance the SBBC strength of $D^0{\bar D}^0$ and $\phi\phi$ in the low momentum region
but suppress the SBBC in the larger momentum region.
For static sources, the squeezing effect does not affect the HBT and the HBT {\color{black}radii} remain almost unchanged with the increasing pair momentum $K_{12}$.
{\color{black}For expanding sources, the HBT radii decrease as the pair momentum $K_{12}$ increases due to the influence of expanding flow}. The squeezing effect suppresses the influence of flow on the HBT {\color{black}radii},
which is significant for two identical bosons with large pair momentum {\color{black}or with large mass}. Due to the squeezing effect, the relationship between the HBT {\color{black}radii} and the pair momentum exhibits non-flow behavior for $D^0{\bar D}^0$ pair. Likewise, non-flow behavior also appears in the HBT {\color{black}radii} of $\phi\phi$ with large pair momentum.

As the collision energy increases, the source medium becomes hotter and denser. Consequently, the bosons may experience a more pronounced medium effect and the time distribution of the source becomes wider. For sources with wide temporal distribution, the SBBC may be suppressed to no signal but the non-flow behavior of the
HBT {\color{black}radii} persist. {\color{black}The results presented in this paper are solely based on theoretical simulations.
Unfortunately, there is currently no SBBC and HBT experimental data available for the $D^0$ meson and the $\phi$ meson for comparative purposes.
The significance of the study lies in the fact that, apart from SBBC, the non-flow behavior of the HBT radii may introduce a novel way to investigate squeezing effects.}

In this paper, the expanding Gaussian source with non-zero width was used to studied the SBBC.
Exploring the SBBC of bosons with non-zero width using more realistic sources, such as the hydrodynamical sources,
would offer a highly intriguing avenue for further investigation.
{\color{black}To study the influence of the squeezing effect on HBT radii, a one-dimensional Gaussian fitting formula is utilized.
However, the spatial distribution of the source may deviate from a Gaussian shape. It is needed to conduct further analysis on the impact of the squeezing effect using fitting formulas that include the form of the source distribution, such as L\'{e}vy-type formula \cite{levy1,levy2,levy3}.}
It would also be interesting to conduct additional research on the influence of the squeezing effect on
the three-dimensional HBT {\color{black}radii}.

\begin{acknowledgements}
This research was supported by the National Natural Science Foundation of China
under Grant Nos. 11905085.
\end{acknowledgements}

\end{document}